\documentclass[aps,twocolumn,superscriptaddress,showkeys,prd,nofootinbib,preprintnumbers]{revtex4-1}

\usepackage{graphicx}
\usepackage{amsmath}
\usepackage{amsfonts}
\usepackage{amssymb}
\usepackage{hyperref}
\usepackage{bm}

\usepackage{MnSymbol}

\def\OO{\mathcal{O}}

\def\Eq#1{Eq.~\eqref{#1}}
\def\nn{\nonumber}
\def\Rep{\text{Re}}
\def\Imp{\text{Im}}
\def \als{\alpha_\mathrm{s}}

\newcommand{\figref}[1]{Fig.~\ref{#1}}						

\newcommand{\CF}{C_{F}}

\newcommand{\Nf}{N_{\text{f}}}
\newcommand{\Nc}{N_{\text{c}}}
\newcommand{\gB}{g_{\text{B}}}
\newcommand{\Gleich}[1]{\stackrel{\text{\makebox[0pt]{$#1$}}}{=}}
\newcommand{\Tr}{\text{Tr}}

\begin{document}

\title{Thermal Heavy Quark Self-Energy from Euclidean Correlators}
\date{\today}

\author{Alexander M. Eller}
\email[]{meller@theorie.ikp.physik.tu-darmstadt.de}
\affiliation{Institut f\"ur Kernphysik, Technische Universit\"at Darmstadt\\
Schlossgartenstra{\ss}e 2, D-64289 Darmstadt, Germany}
\author{Jacopo Ghiglieri}
\email[]{jacopo.ghiglieri@cern.ch}
\affiliation{Theoretical Physics Department, CERN, CH-1211 Gen\`eve 23, Switzerland}
\altaffiliation{Now at SUBATECH UMR 6457 (IMT-Atlantique, Universit\'e de Nantes, CNRS-IN2P3) F-44307 Nantes, France}
\author{Guy D. Moore}
\email[]{guy.moore@physik.tu-darmstadt.de}
\affiliation{Institut f\"ur Kernphysik, Technische Universit\"at Darmstadt\\
Schlossgartenstra{\ss}e 2, D-64289 Darmstadt, Germany}

\preprint{CERN-TH-2019-025}

\begin{abstract}
Brambilla, Escobedo, Soto, and Vairo have derived an effective
description of quarkonium with two parameters; a momentum diffusion
term and a real self-energy term.  We point out that there is a similar real
self-energy term for a single open heavy flavor and that it can be expressed directly in terms of Euclidean
electric field correlators along a Polyakov line.  This quantity can
be directly studied on the lattice without the need for analytical
continuation.  We show that Minkowski-space calculations
of this correlator correspond with the known NLO Euclidean value of
the relevant electric field two-point function and that it differs from the 
real self-energy term for quarkonium.
\end{abstract}

\keywords{quarkonium, quark-gluon plasma, electric--field correlator}

\maketitle
\section{Introduction}
\label{sec:intro}
Quarkonium (bound heavy quark-antiquark states) are an intriguing
probe of the quark-gluon plasma \cite{Andronic:2015wma}.  Originally
proposed by Matsui and Satz \cite{Matsui:1986dk}, the suppression of
quarkonia has remained an active topic of experimental
\cite{Adare:2014hje,Adamczyk:2013tvk,Adamczyk:2013poh,Abelev:2013ila,Adam:2016rdg,Khachatryan:2016xxp}
and theoretical
\cite{Rapp:2017chc,Brambilla:2008cx,Brambilla:2010cs,%
Mocsy:2013syh,Burnier:2015tda,Brambilla:2016wgg,Brambilla:2017zei}
investigation ever since.  The central idea is that a thermal medium
tends to break up quarkonium bound states; one then investigates how
strong this effect is expected to be theoretically, and how much such
states are suppressed experimentally.  Recently it has become clear
that, at the highest collision energies, charmonium experiences important
recombination effects from the many open charm quarks in the plasma
\cite{Young:2009tj,Zhao:2011cv,Blaizot:2015hya}.

Recently Brambilla, Escobedo, Soto, and Vairo have used potential
non-relativistic QCD (pNRQCD) \cite{Pineda:1997bj,Brambilla:1999xf,Brambilla:2004jw}
at second order in the multipole expansion to rigorously
derive in \cite{Brambilla:2016wgg}
 an open quantum system effective description for quarkonium
evolution in a
quark-gluon plasma for $m\gg 1/a_0\gg T$,
where $m$ is the heavy quark mass and $a_0\sim 1/(m\als)$ the Bohr radius.
Their description depends on two (in principle nonperturbative)
parameters describing the interaction of the thermal medium with heavy
quarks.  One parameter is the well-known heavy-quark momentum-diffusion
coefficient \cite{CasalderreySolana:2006rq}
\begin{equation}
  \kappa = \frac{g^2}{6N_c} \mathrm{Re}
  \int_{-\infty}^\infty ds \,\langle \mathrm{T} \:
  E^{a,i}(s,\mathbf{0}) E^{a,i}(0,\mathbf{0}) \rangle \,,
\end{equation}
where $E^{a,i}$ is the color electric field, $s$ is the Minkowski time
value, both $E$ fields are at the same space coordinate $\mathbf{0}$,
$N_c=3$ is the number of colors, $\mathrm{T}$ is
the time ordering symbol, and 
a 
Wilson line is implicitly included in the time-ordered
correlator indicated, i.e., in the notation of \cite{Brambilla:2016wgg,Brambilla:2017zei} 
the  $\vec E$ field has been redefined as $\vec E\to \Omega\vec E \Omega^\dagger$,
with $\Omega\equiv \mathrm{P}\exp\,-i g\int_{-\infty}^t ds A^0(s,\mathbf{0})$.   
This quantity has been extensively investigated
in the literature, both in weak-coupling QCD
\cite{Svetitsky:1987gq,Moore:2004tg,CaronHuot:2007gq,CaronHuot:2008uh},
effective models \cite{vanHees:2007me},
holographic dual theories
\cite{CasalderreySolana:2006rq,Herzog:2006gh},
and via analytical continuation from lattice data
\cite{CaronHuot:2009uh,Ding:2011hr,Francis:2011gc,Banerjee:2011ra,Francis:2015daa}.

The second parameter is a real non-dissipative plasma effect which
induces a mass shift in the heavy-quark bound states.  At lowest order
in pNRQCD, the shift is
$\delta m = \frac{3}{2} a_{0}^{2}\gamma$ \cite{Brambilla:2008cx},
where
 $\gamma$ is the following electric-field correlator
\begin{equation}
\label{eq: mass shift}
\gamma=\frac{g^{2}}{6 \Nc} \Imp \int_{-\infty}^{\infty}ds~\left\langle
\text{T} E^{a,i}(s,\mathbf{0})E^{a,i}(0,\mathbf{0})  \right\rangle \,.
\end{equation}
This correlator has
received less attention in the literature.
In this paper we show that there are two distinct operators, depending
on the Wilson lines connecting the electric fields;
\cite{Brambilla:2008cx}
\begin{equation}
  \label{eq: mass shift adj}
  \gamma_\mathrm{adj}=\frac{g^{2}}{6 \Nc} \Imp \int_{-\infty}^{\infty}ds~\left\langle
  \text{T} E^{a,i}(s,\mathbf{0})U(s,0)_{ab}E^{b,i}(0,\mathbf{0})  \right\rangle \,,
  \end{equation}
with $U(s,0)_{ab}$ an adjoint-representation%
\footnote{%
  To fix clearly the conventions, we choose
  $D_\mu=\partial_\mu-ig A_\mu$, so that
  $U(a,b)\equiv \mathrm{P}\exp\,i g\int_b^a ds A^0(s,\mathbf{0})$
  is the Wilson line in either representation.}
Wilson line, and 
\begin{align}
  \label{eq: mass shift fund}
  \gamma&_{\mathrm{fund}}=\frac{g^{2}}{3 \Nc} \Imp \int_{-\infty}^{\infty}ds~
  \\
  \nonumber
  \times& \left\langle \Tr\left[
  \text{P} ~U(-\infty,s) E^{i}(s,\mathbf{0}) U(s,0) E^{i}(0,\mathbf{0}) U(0,-\infty) \right]  \right\rangle, 
  \end{align}
with $E^i=E^{a,i} T^a$, the Wilson lines $U$ in the fundamental
representation and P is the path-ordering symbol, indicating that fields
are contour-ordered along the Wilson lines; this equals time ordering for the $E$ fields 
but not for the $U$ operators.

The different Wilson line structures reflect the different time evolution of the two objects.
The former corresponds to the real part of the heavy quarkonium singlet self-energy.
Before (after) the pNRQCD dipole vertex at $t=0$ ($t=s$) the bound state is a color singlet
and thus only the adjoint Wilson line connecting the two $E$ fields exists in Eq.\eqref{eq: mass shift adj}.
Conversely, Eq.~\eqref{eq: mass shift fund} is linked to the real part
of the open heavy-quark self-energy:
the Wilson lines are always fundamental and reflect the fact that the heavy quark can  interact
with the environment at all times.

In this paper we show that $\gamma_\mathrm{adj}$, which was calculated to LO in Ref.~\cite{Brambilla:2008cx},
differs from $\gamma_\mathrm{fund}$, whose LO computation we present here for the first time. 
We also show that the latter object can be  
determined via lattice QCD much more easily than
the coefficient $\kappa$, as it maps directly to a Euclidean expression,
without the need of analytic continuation or spectral function reconstruction.  
Therefore we will focus in this paper on
 the coefficient $\gamma_{\mathrm{fund}}$. We plan to investigate the
Euclidean mapping of $\gamma_{\mathrm{adj}}$ in a follow-up paper~\cite{next}, together with
the physical interpretation of $\gamma_{\mathrm{fund}}$.
This follow-up paper will also discuss the difference between
$\kappa_{\mathrm{adj}}$ and $\kappa_{\mathrm{fund}}$, which are also
distinct objects.

In the next section we will show how to analytically continue
Eq.~(\ref{eq: mass shift fund}) to Euclidean time.  This leads to a
time-integral of a correlator of electric fields along a Polyakov
loop.  In the remainder of the paper we check this derivation by
showing that the next-to-leading order (NLO) value of $\gamma_{\mathrm{fund}}$, derived by Minkowski
space methods, correctly corresponds to the appropriate
integral moment of the known NLO Euclidean correlator from
Ref.~\cite{Burnier:2010rp}.  This is a nice check of our continuation.

\section{Analytic Continuation of Electric-Field Correlator}
\label{sec:formal}
The analytic continuation of the electric-field correlator Eq.~\eqref{eq: mass shift fund} 
is not trivially doable. However, we don't consider this correlator directly, 
but instead use the heavy-quark current-current correlator
\begin{align}
\label{eq: current current correlator}
\int_{-\infty}^{\infty}dt ~e^{i\omega t} \int d^{3}\mathbf{x}~\left\langle \left[ \mathcal{\hat{J}}^{\mu}(t,\mathbf{x}) , \mathcal{\hat{J}}^{\nu}(0,\mathbf{0}) \right]  \right\rangle\,,
\end{align}
which Eq.~\eqref{eq: mass shift fund} originates from. Here
$\mathcal{\hat{J}}^{\mu}\equiv \hat{\bar{\psi}} \gamma^\mu \hat{\psi}$
is the heavy-quark current and $\hat{\psi}$ is the heavy-quark field
operator. The definition of the heavy-quark momentum-diffusion
coefficient $\kappa$ as the spectral function of Eq.~\eqref{eq:
  current current correlator} in the $M_{\text{kin}}\rightarrow
\infty$ limit is derived in Ref.~\cite{CaronHuot:2009uh}.
Its authors also show that in the heavy-quark limit the current-current correlator can be analytically continued to Euclidean time leading to
the Euclidean color--electric correlator
\begin{align}
\label{eq: color--electric correlator}
G&_{\text{E}}^{\text{HQ}}(\tau)
\\ \nn
&= -\!\sum_{i=1}^{3-2\epsilon} \frac{\left\langle \Rep~ \Tr \left[ U(\beta,\tau)\gB E_{i}(\tau,\mathbf{0})U(\tau,0) \gB E_{i}(0,\mathbf{0}) \right] \right\rangle}{3\left\langle \Rep~ \Tr \left[U(\beta,0) \right] \right\rangle } \,,
\end{align}
where the minus sign  emerges 
from different factors of $i$ in the definition of the color--electric field in 
real and imaginary time.
For the details of the continuation we refer the reader to the original paper \cite{CaronHuot:2009uh}.

In order to analytically continue $\gamma_{\mathrm{fund}}$, we need to relate the imaginary part of Eq.~\eqref{eq: current current correlator} with its analytic continuation Eq.~\eqref{eq: color--electric correlator}. Therefore we want to remind the reader of the relation between the imaginary part of a two-point function of two hermitian operators $A,B$ in real time with the zero Matsubara frequency limit of the corresponding Euclidean correlator, such that
\begin{align}
\begin{split}
\label{eq: im of time integrated correlator}
\Imp~ G_{R}^{AB}(\omega{=}0) = {} & \Imp \int_{0}^{\infty}dt~ G^{AB}(t)
\\
{} = {} & -\int_{0}^{\beta} d\tau~G_{E}^{AB}(\tau)= -\tilde{G}_{E}(\omega_n{=}0)\,.
\end{split}
\end{align}
The two-point functions are defined in the usual way, $G^{AB}(t)=
\Tr\left\lbrace \hat{\rho} \left[A(t),B(0) \right] \right\rbrace$ and
$G^{AB}_{E}(\tau)= \Tr\left\lbrace \hat{\rho} A(-i\tau)B(0)
\right\rbrace$  with $\hat{\rho}\equiv \frac{1}{Z}e^{-\beta H}$ the
finite-temperature equilibrium density matrix.
(Note that $\int_0^\infty dt G^{AB}(t)$ is purely imaginary, since the
commutator of Hermitian operators gives twice the imaginary part.
Nevertheless, we take the imaginary part explicitly because the
finite-frequency transform contains real and imaginary parts.)

We insert two complete sets of energy eigenstates in the definition of
the two-point function such that the LHS of Eq.~(\ref{eq: im of time
  integrated correlator}) becomes
\begin{align*}
&\Imp \int_{0}^{\infty}dt~ \sum_{n,m} \frac{2}{Z}A_{mn} B_{nm}
 e^{-\frac{\beta}{2}(E_{n}+E_{m})}
\\
& \qquad \times \sinh\left( \frac{\beta (E_{n}-E_{m})}{2} \right) e^{-i(E_{n}-E_{m})t}
\\
&=-\sum_{n,m} \frac{2}{Z} \mathrm{Re}\left[A_{mn} B_{nm}\right] \frac{e^{-\frac{\beta}{2}(E_{n}+E_{m})}}{E_{n}-E_{m}} \sinh\left( \frac{\beta (E_{n}-E_{m})}{2} \right)\,,
\end{align*}
where we used the notation $A_{nm}=\left\langle n \left| A(t=0) \right|m \right\rangle$.\\
Using the same procedure on the RHS of Eq.~(\ref{eq: im of time integrated correlator}) yields
\begin{align*}
&\int_{0}^{\beta}d\tau~ \sum_{n,m} \frac{1}{Z} A_{mn} B_{nm} e^{-\beta E_{m}} e^{-(E_{n}-E_{m})\tau}
\\
&=\sum_{n,m} \frac{2}{Z} A_{mn} B_{nm} \frac{e^{-\frac{\beta}{2}(E_{n}+E_{m})}}{E_{n}-E_{m}} \sinh\left( \frac{\beta (E_{n}-E_{m})}{2} \right)\,.
\end{align*}
So as long as $A_{mn} B_{nm}$ has no imaginary part, which happens
if $A$ and $B$ are Hermitian operators or if $A=B^\dagger$,
 Eq.~(\ref{eq: im of time integrated correlator}) is true. 
From this we conclude that the analytic continuation of $\gamma_{\mathrm{fund}}$ is
\begin{equation}
  \label{gammaeuclid}
\gamma_{\mathrm{fund}}=-\int_{0}^{\beta}d\tau~  G_{\text{E}}^{\text{HQ}}(\tau)\,,
\end{equation}
One of the main results of this paper is therefore that the thermal effects on 
 $\gamma_{\mathrm{fund}}$ can be determined by a nonperturbative calculation using the 
vacuum-subtracted Euclidean color--electric correlator on the lattice.

To further clarify the need for vacuum subtraction,
let us look at the Euclidean color--electric correlator
at leading order (LO), $\OO(g^2)$. It is obtained trivially
by connecting the two chromoelectric fields with a gluon propagator, yielding
\cite{CaronHuot:2009uh,Burnier:2010rp}
\begin{equation}
  \label{lomatsu}
  G_\mathrm{E\,LO}^\mathrm{HQ}(\tau)=-\frac{g^2C_F}{3}\sumint\limits_{K}e^{i k_n\tau}\frac{(D-1)k_n^2+k^2}
  {k_n^2+k^2},
\end{equation}
where $\sumint\limits_{K}\equiv T\sum_{k_n}\int_k $, $\int_k \equiv \int d^dk/(2\pi)^d$ with $D=d+1$ the
dimension of spacetime and $k_n$ the bosonic Matsubara frequency. One could
immediately perform the $\tau$ integration of Eq.~\eqref{gammaeuclid},
obtaining $\beta\delta_{k_n}$, at which point the $\int_k$ integral would vanish
 in dimensional regularization (DR). So would, in this scheme, the vacuum contribution,
 where the $\tau$ integrations runs from $-\infty$ to $+\infty$ and the
 Matsubara sum is replaced with an integral over a continuous
 Euclidean frequency $k_4$.
  However, to better illustrate the need for
vacuum subtraction in other schemes, such as the lattice,
let us instead perform first the Matsubara sum and then the $\int_k$, which gives
\cite{CaronHuot:2009uh}
\begin{equation}
  \label{lomatsusum}
  G_\mathrm{E\,LO}^\mathrm{HQ}(\tau)=g^2 C_F \pi^2 T^4\left[\frac{\cos^2(\pi\tau T)}
  {\sin^4(\pi\tau T)}+\frac{1}{3\sin^2(\pi\tau T)}\right].
\end{equation}
The integration of this object over the compactified time direction does not
converge, as the integrand diverges as $\tau^{-4}$ as $\tau \to 0$ and
as $(\beta - \tau)^{-4}$ as $\tau \to \beta$.
But this divergence is ultraviolet, as it comes about when the two $E$ fields
are brought together. It is thus equal to the behavior observed in vacuum,
which can be easily obtained from the $k_4$ integration, leading to
\begin{equation}
  \label{vacuum}
  G_\mathrm{E\,LO}^\mathrm{HQ}(\tau,T=0)=\frac{g^2 C_F}{ \pi^2\tau^4}.
\end{equation}
Hence, vacuum subtraction in a non-DR scheme takes the form
\begin{align}
  \label{vacsubtr}
  \gamma_{\mathrm{fund}}=&-2\int_0^{\beta/2}d\tau\left[G_\mathrm{E\,LO}^\mathrm{HQ}(\tau)
  -G_\mathrm{E\,LO}^\mathrm{HQ}(\tau,T=0)\right]\nn\\
  &+2\int_{\beta/2}^\infty d\tau\,
  G_\mathrm{E\,LO}^\mathrm{HQ}(\tau,T=0)=0+\OO(g^4),
\end{align}
where we have exploited the symmetry of the thermal contribution at
$\tau=\beta/2$ and that of the vacuum at $\tau=0$. It is precisely
a subtraction of this kind that would need to be performed on the lattice:
for all $\tau<\beta/2$ values, one computes the difference between the
correlator on the thermal lattice and the vacuum lattice, and one then
subtracts the integral of the vacuum contribution over $\tau>\beta/2$.
In practice, due to the noisy denominator in
\Eq{eq: color--electric correlator}, it may be impossible to subtract
$\gamma_{\mathrm{fund}}$ at very low temperature on the lattice; in practice a
subtraction at a temperature where thermal effects are expected to be
small should be sufficient.

At small separation, where the vacuum and thermal correlators diverge
but the difference stays finite, it may be difficult to extract the
difference with good statistical power.  However, we believe that,
while the individual short-distance values are sensitive to even small
amounts of gradient flow
\cite{Narayanan:2006rf,Luscher:2009eq,Luscher:2010iy},
the difference should not be.  This is supported by existing
analytical studies \cite{Eller:2018yje}, and it would be useful to
investigate this issue further. We also refer to \cite{Christensen:2016wdo}
for the perturbative $\OO(\als)$ renormalization of
Eq.~\eqref{eq: color--electric correlator} in the lattice scheme.

\section{NLO Integration of the Euclidean correlator}
\label{sec:diagrams}
\begin{figure}[tb]
  \hfill \includegraphics[width = \columnwidth]{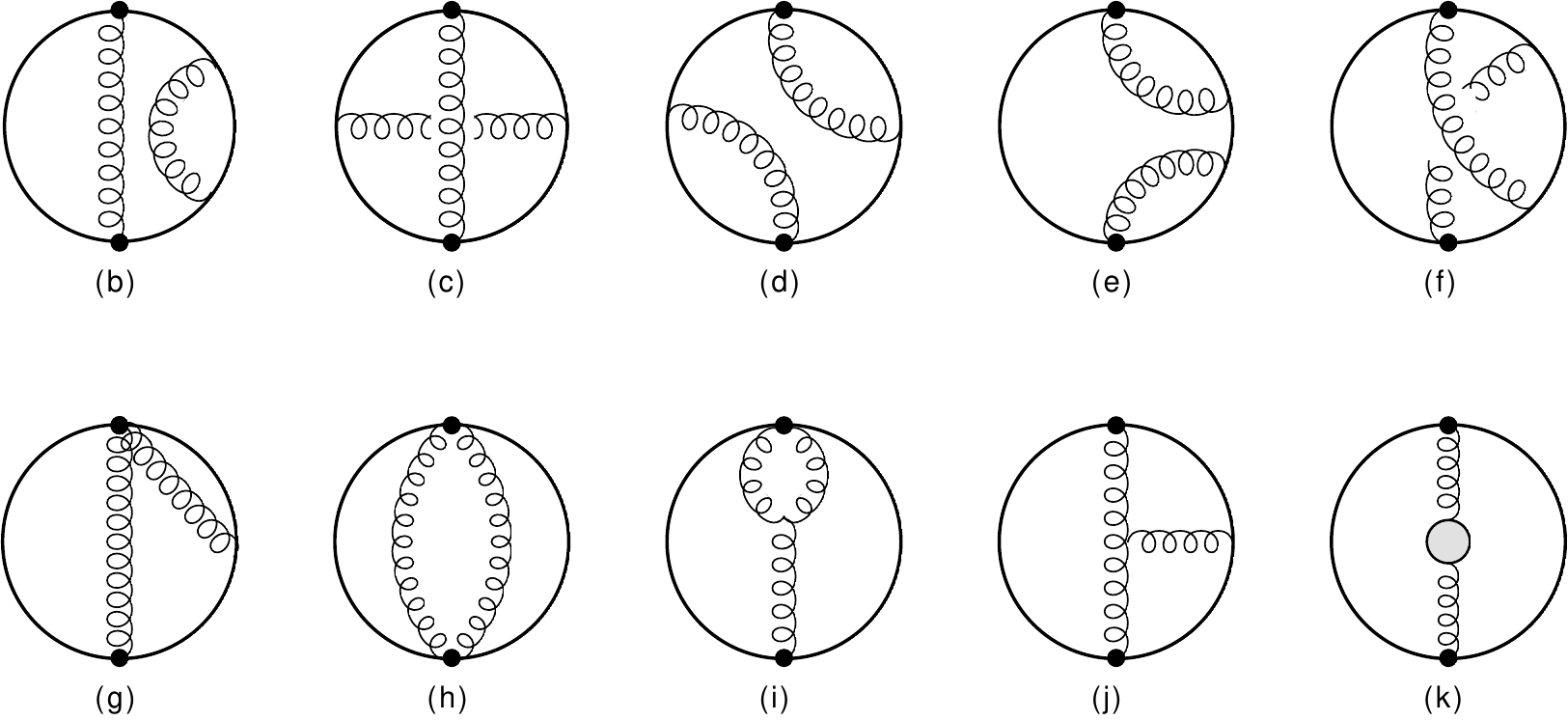}
  \hfill $\phantom{.}$
  \caption{\label{fig:g4 feynman diagrams}
Feynman diagrams contributing to the correlation function at order
$g^{4}$. The circle represents the Polyakov line and the heavy dots
are the electric field insertions.  Diagram $(k)$ includes the full
gluon self energy and therefore fermionic and ghost contributions.
The labeling of the diagrams parallels that of Ref.~\cite{Burnier:2010rp}; 
diagram $(a)$, not shown, is the order-$g^2$ graph.}
\end{figure}

In this section, we validate our result in a perturbative calculation at 
next--to--leading order: we first 
present the imaginary-time integration of the correlator in Eq.~\eqref{eq: color--electric correlator}, 
to be followed by the corresponding real-time counterpart. 
The Euclidean color--electric correlator was calculated in perturbation theory up to next--to--leading order in Ref.~\cite{Burnier:2010rp} and the contributing diagrams are shown in \figref{fig:g4 feynman diagrams}.
Since we are interested in the thermal contributions to $\gamma_{\mathrm{fund}}$, we subtract the vacuum contribution of the correlator during the calculation. As previously highlighted,
dimensional regularization does that automatically, so in the following
we will not keep track of scale-free contributions that vanish in any $D$.

Using the integral expression of $G_\mathrm{E\,NLO}^\mathrm{HQ}(\tau)$
obtained in Ref.~\citep{Burnier:2010rp} from the diagrams above,
 we notice that the only $\tau$ dependence is in the Fourier transform.
After applying the  Kronecker delta $\beta \delta_{k_{n}}$ arising
from the $\tau$ integration,
we obtain that\footnote{We thank Viljami Leino for the discovery of an issue in our
evaluation of Eq.~\eqref{mastint}, then reflected in Eq.~\eqref{loresult}, 
in a previous version of this article. We also thank M.~\'A.~Escobedo, N.~Brambilla,
J.~Soto and A.~Vairo for comments on this revised version.}
\begin{align}
\nn  \gamma_{\mathrm{fund}}^\mathrm{LO}=&-\int_0^\beta d\tau\, G_\mathrm{E\,NLO}^\mathrm{HQ}(\tau)\\
  =&
  -\frac{g^4C_F}{3}N_f(\tilde{\mathcal{I}}_1+4\,\tilde{\mathcal{I}}_2),
  \label{mastint}
\end{align}
where we have introduced these sum integrals
\begin{align}
\label{eq: k sum int part1}
\mathcal{I}_{1}\,,\,\tilde{\mathcal{I}}_{1}=
&\int_{k} \sumint\limits_{Q, \left\lbrace Q \right\rbrace}  \frac{1}{Q^{2} (K-Q)^{2}}\bigg\vert_{k_n=0} \, ,
\\
\label{eq: k sum int part2}
\mathcal{I}_{2} \,,\,\tilde{\mathcal{I}}_{2}     =
&\int_{k} \sumint\limits_{Q, \left\lbrace Q \right\rbrace}  \frac{q_{n}^{2}}{K^{2}Q^{2} (K-Q)^{2}} \bigg\vert_{k_n=0}\, ,
\end{align}
where $K^{2}=k_{n}^{2}+ \mathbf{k}^{2}$.
The notation $\left\lbrace Q \right\rbrace$ represents the fermionic Matsubara frequencies $q_{n}=\pi T(2n+1)$ for the fermionic sum integrals, denoted
by $\tilde{\mathcal{I}}$. These are directly related to the bosonic ones via
\begin{align}
\label{eq: relation fermionic and bosinic summs}
\sigma_{f}(T)&=T \sum_{\left\lbrace q_{n} \right\rbrace} f(q_{n},\dots)= 2 \sigma_{b}\left(T/2 \right) -  \sigma_{b}(T)  \,.
\end{align}

The bosonic integrals are easily evaluated as
\begin{align}
  \label{i1}
  \mathcal{I}_1=&\frac{\Gamma^2 (1-d/2)\zeta (4-2 d)}
  {8\,\pi ^{4-d}\, T^{3-2 d}} ~~~ \Gleich{d \rightarrow 3} ~~~0\,,\\
  \label{i2}
  \mathcal{I}_2=&-\frac{(d-2)   \Gamma^2 (1-d/2)
  \zeta (4-2 d) }{16 (d-3)\pi ^{4-d}\,T^{3-2d}}~~ \Gleich{d \rightarrow 3} ~~-\frac{\zeta(3)}{8\pi^2\beta^3},
\end{align}
where we have used a Feynman parameter for $\mathcal{I}_2$.
The fermionic counterparts, obtained from
 Eq.~\eqref{eq: relation fermionic and bosinic summs}, are
\begin{align}
\nn
\mathcal{I}_{a} \propto \frac{1}{\beta^{2d-3}} \rightarrow \tilde{\mathcal{I}}_{a} = \mathcal{I}_{a}  \times (4^{2-d}-1) \,.
\end{align}
Putting everything together, we find
\begin{equation}
  \label{loresult}
\gamma_{\mathrm{fund}}^\mathrm{LO} =-2 \als^{2} T^{3} \zeta(3) \CF \Nf   \,.
\end{equation}
This results differs from the real-time calculation of Eq.~\eqref{eq: mass shift adj} with its 
adjoint Wilson line given in Ref.~\cite{Brambilla:2008cx}, which found
\begin{equation}
  \label{loresultadj}
\gamma_{\mathrm{adj}}^\mathrm{LO} =-2 \als^{2} T^{3} \zeta(3) \CF\left(\frac43 \Nc+ \Nf\right)  .
\end{equation}
To our knowledge there is no real-time determination of Eq.~\eqref{eq: mass shift fund} 
in the literature; let us thus present it briefly. We find Coulomb gauge to be a good choice:
in this gauge the $A_0A_0$ retarded bare propagator is $G_{00}^R(q^0,q)=i/q^2$ and equals 
the advanced one. Thus, the spectral density vanishes, making the off-diagonal entries
of the propagator matrix in the ``12'' formalism of real-time perturbation theory vanish.
Moreover, the lack of frequency dependence of the diagonal elements
makes many diagrams vanish in DR. We label the real-time graphs as in Fig.~\ref{fig:g4 feynman diagrams},
though the Wilson lines now start and end at $t=-\infty$. The only non-vanishing diagrams
in this gauge are then (i), (j) and (k). Of these, (i) and (k) do not source any gluons
from the Wilson lines: they thus contribute equally to $\gamma_\mathrm{adj}$ and 
$\gamma_\mathrm{fund}$, as the color trace gives the same result. In this gauge, any difference between the two can thus only arise
from diagram (j) and its equivalent for $\gamma_\mathrm{adj}$.  We 
  show in App.~\ref{app_cg} how the Coulomb gauge evaluation of the contribution
of this diagram to Eqs.~\eqref{eq: mass shift adj} and \eqref{eq: mass shift fund} yields
the difference between Eqs.~\eqref{loresultadj} and \eqref{loresult},\footnote{Ref.~\cite{Brambilla:2008cx}
obtained Eq.~\eqref{loresultadj} in the temporal axial gauge $A_0=0$. We have also checked that
the sum of diagrams (i), (j) and (k) in the expansion of $\gamma_\mathrm{adj}$ in Coulomb gauge reproduces Eq.~\eqref{loresultadj}.}
thus confirming the correctness of our analytical continuation, Eq.~\eqref{gammaeuclid}, to first non-trivial
order in perturbation theory.

Interestingly, it is also possible to go beyond
this order. 
Our Euclidean analysis has so far used unresummed perturbation theory, which
is appropriate when all momenta
are of order $T$ and the Matsubara frequency $k_n$ is nonzero.  Unlike
$\kappa$, which receives a contribution from the $gT$ scale at LO, $\gamma_{\mathrm{fund}}$
does not. Inspection of Eqs.~\eqref{gammaeuclid} and
\eqref{lomatsu} shows that $k\sim gT$ contributes
to $\gamma_{\mathrm{fund}}$ at $\OO(g^5)$. This contribution is easily obtained
by replacing Eq.~\eqref{lomatsu} with its resummed version. Since the $\tau$
integration forces $k_n=0$, it suffices to use Electrostatic QCD (EQCD)
\cite{Braaten:1994na,Braaten:1995cm,Braaten:1995jr,Kajantie:1995dw,Kajantie:1997tt},
where ${\bf E}\approx-i \boldsymbol{\nabla} A_0$.
Then the temporal component of the gauge field gets Debye-screened,
yielding
\begin{equation}
  \label{eqcd}
\gamma_\mathrm{NLO}
=\frac{g^2C_F}{3}\int_k\frac{k^2}
{k^2+m_D^2}=\frac{\als C_Fm_D^3}{3},
\end{equation}
where $m_D^2=g^2 T^2(\Nc/3+\Nf/6)$ is the Debye mass. 
Eq.~\eqref{eqcd} agrees with Eq.~(87) of \cite{Brambilla:2008cx}, recalling
that $\mathrm{Re}\,\delta V_s(r)_{11}^{\text{\cite{Brambilla:2008cx}}}=\gamma_\mathrm{adj} \,r^2/2$. 
As the Wilson lines do not contribute,
this $\OO(g^5)$ contribution $\gamma_\mathrm{NLO}$ is identical for $\gamma_\mathrm{fund}$ and $\gamma_\mathrm{adj}$ and is  the
only NLO contribution to both $\gamma$; to the best of our knowledge, this was
not observed in the previous literature.

\section{Discussion and conclusions}
\label{discussion}
In this paper we considered  the coefficient $\gamma_{\mathrm{adj}}$, introduced by
Brambilla, Escobedo, Soto and Vairo \cite{Brambilla:2016wgg} and defined in Eq.~\eqref{eq: mass shift adj}, which
describes the in-medium mass shift to heavy quarkonium for $m\gg 1/a_0\gg T$, and
 $\gamma_{\mathrm{fund}}$ (Eq.~\eqref{eq: mass shift fund}). The latter object
can be seen as the fluctuation counterpart to $\kappa$, the momentum-diffusion
coefficient for a single heavy quark. We have shown that
$\gamma_{\mathrm{fund}}$ and $\gamma_{\mathrm{adj}}$ are not equal; 
this was done by presenting an explicit calculation
for $\gamma_{\mathrm{fund}}$, which shows how it differs from $\gamma_{\mathrm{adj}}$ at 
the first non-trivial order in perturbation theory, $\mathcal{O}(g^4T^3)$. 
Physically, the difference between the two can be understood as follows: $\gamma_{\mathrm{fund}}$
is related to the propagation of a single heavy quark, which can interact with the medium at any time,
whereas  $\gamma_{\mathrm{adj}}$ describes a $Q\bar Q$ pair, which is a medium-blind singlet
before (after) the first (last) $E$ field insertion. Indeed, our explicit evaluation, presented in
App.~\ref{app_cg}, shows how the difference arises from the Wilson lines before/after the $E$ fields.\footnote{
  \label{foot_tag}
  The computation of $\gamma_{\mathrm{adj}}$  in \cite{Brambilla:2008cx} was performed in the $A^0=0$ gauge, where
  the contribution of the Wilson lines vanishes and one would naively expect $\gamma_{\mathrm{adj}}$ and 
  $\gamma_{\mathrm{fund}}$ to be equal. However, more care is needed when temporal Wilson lines
  stretch to $t=-\infty$ in the $A^0=0$ gauge; we plan to return to the issue of gauge invariance in this singular
  gauge in \cite{next}.}
We also observe that, at zero temperature, the NLO result for $g^2/(6\Nc)\langle E^a(t)U_{ab}(t,0)E^b(0)\rangle$ 
in~\cite{Eidemuller:1997bb}
and that for  $g^2/(3\Nc)\left\langle \Tr\left[
  \text{T} ~U(-\infty,t) E^{i}(t) U(t,0) E^{i}(0) U(0,-\infty) \right]  \right\rangle$, which can be extracted
  from the $T=0$ limit of~\cite{Burnier:2010rp}, do not agree, further confirming the different nature
of the two operators.

We have  furthermore shown that $\gamma_{\mathrm{fund}}$
can be re-expressed in terms of a Euclidean correlation
function, \eqref{eq: color--electric correlator}, which is highly
amenable to a lattice determination.  With the vacuum contributions
removed, the time integral of the correlator, \eqref{gammaeuclid},
 should not suffer from divergences and the
computational cost should be reasonable if smoothing techniques like
gradient flow are employed. We confirmed that the LO results for
$\gamma_{\mathrm{fund}}$, evaluated via real-time techniques, agree with the Euclidean
time-integration of the results of Ref.~\cite{Burnier:2010rp}, which
is a nontrivial check on our derivation of the Euclidean continuation. We also obtained the NLO
correction to $\gamma_{\mathrm{fund}}$ and $\gamma_{\mathrm{adj}}$ in Eq.~\eqref{eqcd}.

The physical interpretation of $\gamma_{\mathrm{fund}}$
is at the moment however not completely clear to us; we plan to return to this issue in
a follow-up publication, \cite{next}, where we  also intend to address the issue of the Euclidean
counterpart to $\gamma_{\mathrm{adj}}$. Similarly, we can define
$\kappa_{\mathrm{fund}}$
as the real rather than imaginary part of
Eq.~\eqref{eq: mass shift fund},
and $\kappa_{\mathrm{adj}}$ as the real part of
Eq.~\eqref{eq: mass shift adj}.  The former is relevant for the medium
interactions of open heavy quarks, while the latter is relevant in
quarkonium physics.
Perturbative results show that they agree up to order $g^5T^3$
\cite{CaronHuot:2007gq,CaronHuot:2008uh,Brambilla:2008cx}, but there
is no reason why this should persist to all orders.
We plan to touch this issue as well in \cite{next}, together with that
of gauge invariance discussed in footnote~\ref{foot_tag}.

\begin{acknowledgments}
We thank the Technische Universit\"at Darmstadt and its Institut f\"ur Kernphysik,
 where this work was conducted and where JG was hosted during the early phase of this work.
This work was funded by the Deutsche Forschungsgemeinschaft (DFG, German Research Foundation)
 – Project number 315477589 – TRR 211.

\end{acknowledgments}

\section*{Note added}
As we were finalizing this paper, we became aware of the preprint ``Transport
coefficients from in medium quarkonium dynamics'' \cite{munich} by
N.~Brambilla, M.~\'A.~Escobedo, A.~Vairo and P.~Vander Griend. It
proposes a way to determine $\gamma$ from the quarkonium
spectral function reconstructed from lattice QCD \cite{Kim:2018yhk}.
We thank the authors for sharing their results with us prior to publication
and for discussion.
\appendix
\section{Explicit real-time computation of diagram (j)\label{app_cg}}
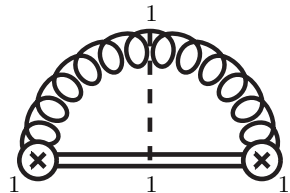
\begin{figure}[ht]
  \begin{center}
    \includegraphics[width=4cm]{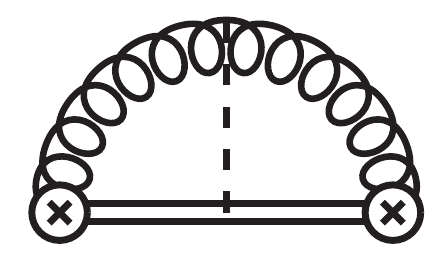}
    \put(-8,0){1}
    \put(-58,0){1}
    \put(-58,65){1}
    \put(-110,0){1}
  \end{center}
  \caption{\label{fig_adjoint}Diagram (j) for $\gamma^{(\mathrm{j})}_\mathrm{adj}$. The vertices with the cross 
  are the $E$ fields, the double line is the adjoint Wilson line, curly lines are transverse gluons and the
  dashed line a temporal gluon. The diagram where one of the $E$ fields sources a temporal gluon is not shown
  explicitly. The ``1'' label the ``12'' assignments of the fields.}
\end{figure}
In the adjoint case, diagram (j) is shown in Fig.~\ref{fig_adjoint}. It contributes to
\begin{align}
  \gamma^{(\mathrm{j})}_\mathrm{adj}
  =& - \frac{g^4}{3 \Nc}  \Imp \int_{0}^{\infty}dt  \int_{0}^{t}dt' 
  \int_Q\int_P e^{i q^0 t}e^{-i (p^0+q^0) t'} \nonumber\\
  &\times\frac{ i f^{acb}  f^{abc}}{(\vec p+\vec q)^2}\bigg[
   q^0 p^0(q^0-p^0)G^{11}_{ik}(P)G^{11}_{ki}(Q)
  \nonumber\\
  &\hspace{0.5cm}+2i p^0\hat{q}^i\hat{q}^jG^{11}_{ij}(P)-2i q^0\hat{p}^i\hat{p}^jG^{11}_{ij}(Q)
  \bigg],
  \label{adjstart}
\end{align}
where we have rewritten the integral over negative and positive times of the contour-ordered operator
as twice the positive-time integral of the forward Wightman operator. Thus, the three fields sourced
by the operator, $E(t)$, $A^0(t')$ and $E(0)$, are naturally time-ordered and thus of type ``1''
in the ``12'' formalism of real-time perturbation theory. In Coulomb gauge the $A^0(t')$ field can only connect to another
$A^0$ field, which has furthermore to be of type ``1'' as well, due to the diagonal nature of the 
bare temporal propagator matrix. Hence, the three-gluon vertex has to be of type ``1'', so that
the propagators of the transverse gluons have to be of type ``11'', i.e., time-ordered. Indeed,
the second line of Eq.~\eqref{adjstart} is the contribution with two transverse gluons sourced
by the two $E$ fields, as depicted in Fig.~\ref{fig_adjoint}, while on the final line they source one transverse and one temporal gluon.
Finally, $\int_P=\int d^DP/(2\pi)^D$
is the Minkowski $D$-dimensional integral.  

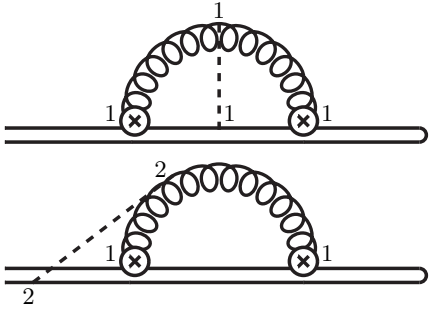
\begin{figure}[ht]
  \begin{center}
    \includegraphics[width=6cm]{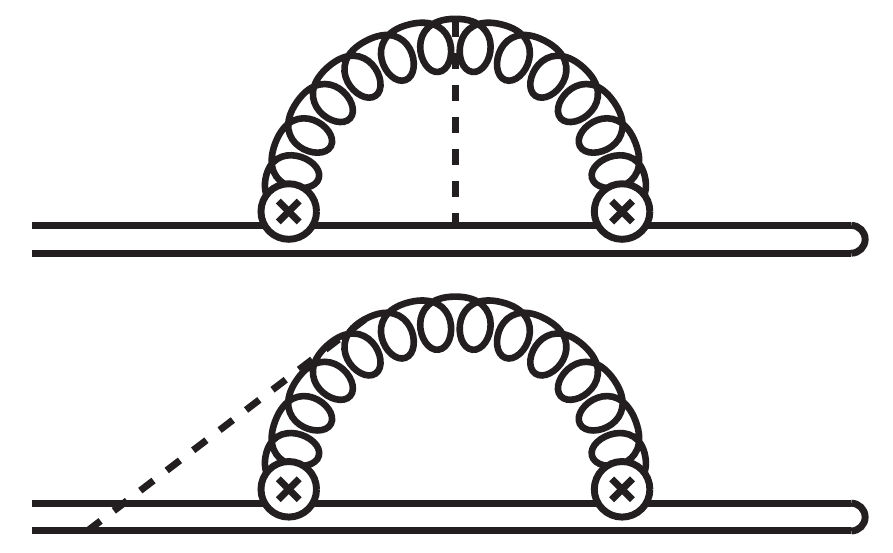}
    \put(-108,44){2}
    \put(-45,12){1}
    \put(-127,12){1}
    \put(-158,-4){2}
    \put(-82,65){1}
    \put(-86,104){1}
    \put(-45,65){1}
    \put(-127,65){1}
  \end{center}
  \caption{\label{fig_fund}Diagrams (j) for $\gamma^{(\mathrm{j})}_\mathrm{fund}$. The graphical notation is the same
  as in Fig.~\ref{fig_adjoint}, except that the solid line is now the Wilson line stretching forward in time from negative to positive
  infinity  passing both $E$ fields (upper contour), to then turn back and return to $-\infty$ (lower contour). We show two of the six
  possibilities for the temporal gluon, which can connect as a ``1'' (``2'') field to the upper  (lower) contour before, between or after
  the $E$ fields. We do not show the case where the $E$ fields source one transverse and one temporal gluon; there the lower contour
  does not contribute.}
\end{figure}
In the fundamental case one has instead the configurations shown in Fig.~\ref{fig_fund}. They give
\begin{align}
  \gamma^{(\mathrm{j})}_\mathrm{fund}
  =&  -\frac{g^{4}}{6 \Nc}  \Imp \int_{0}^{\infty}dt \left(\int_{0}^{t}dt'-\int_{-\infty}^{0}dt'-\int_{t}^{\infty}dt'
  \right) 
  \nn\\
  &\times\int_Q\int_P e^{i q^0 t}e^{-i (p^0+q^0) t'} \frac{ i  f^{acb} f^{abc}}{(\vec p+\vec q)^2}\bigg[
  q^0 p^0(q^0-p^0)\nn\\
 & \times(G^{11}_{ik}(P)G^{\mathrm{11}}_{ki}(Q)+\epsilon(t')\epsilon(t-t')G^{>}_{ik}(P)G^{>}_{ki}(Q))
  \nn\\
  &\hspace{-0.3cm}+ 2ip^0\hat{q}^i\hat{q}^jG^{11}_{ij}(P)- 2i q^0\hat{p}^i\hat{p}^jG^{11}_{ij}(Q)
  \bigg],
  \label{fundstart}
\end{align}
where, as shown in the figure, there are now two possible ``12'' assignments  for the fields sourced by the operator: 
the $E$ fields are always of type ``1'', while the $A_0$ gluon is ``1'' if it comes from $U(0,-\infty)$
or $U(s,0)$, ``2'' if from $U(-\infty,s)$. For reasons which will become clearer soon, 
we have  rewritten this last Wilson line as
 $U(-\infty,s)=U(-\infty,\infty)U(\infty,s)$, with $U(-\infty,\infty)$ thus of type ``2'' and 
 $U(\infty,s)$ of type ``1'', hence the $\int_{t}^{\infty}dt'$ contribution.\footnote{In a covariant gauge the
 contribution of $\int_{t}^{\infty}dt'$ vanishes, as expected from the unitarity of the Wilson lines.
 In Coulomb gauge one needs anyway to consider $U(-\infty,s)$ as $U(-\infty,s+\delta^+)U(\delta^+,s)$, with $\delta^+$
 arbitrarily small and positive. This avoids the appearance of ill-defined $\theta(0)$ contributions
 arising from the time integrations of the bare temporal propagators, which are instantaneous in time.}
 The second line describes the diagrams with two transverse gluons, where we have used the definition $G^>=G^{21}$.
 For these diagrams, as shown in Fig.~\ref{fig_fund}, we
 have two assignments contributing to each of the three $dt'$ integrations.
 The relative sign between the two, encoded in the sign functions $\epsilon(t')\epsilon(t-t')$,
 arises from the combination  of a minus sign from the different color ordering --- in the $0<t'<t$ region ---
  together with another minus sign from the opposite direction of the Wilson lines. The final line encodes the
  contribution of graphs with a single transverse gluon, for which the lower ``2'' contour does not contribute.
  The overall factor of $1/2$ in front of Eq.~\eqref{fundstart}
 with respect to Eq.~\eqref{adjstart} arises from color tracing in the different cases.

 If we take the difference between Eq.~\eqref{fundstart} and \eqref{adjstart} we obtain
  $\Delta\gamma\equiv\gamma_\mathrm{fund}-\gamma_\mathrm{adj}$. It reads   
 \begin{align}
  \Delta\gamma
  =&-  \frac{g^{4}\CF\Nc}{3}  \Imp \int_{0}^{\infty}dt \int_{-\infty}^{\infty}dt'
  \int_Q\int_P
  \frac{ i e^{i q^0 t}e^{-i (p^0+q^0) t'}}{(\vec p+\vec q)^2} \nn\\
  &\times  \bigg\{
  q^0 p^0(q^0-p^0)
 [G^{11}_{ik}(P)G^{\mathrm{11}}_{ki}(Q)-G^{>}_{ik}(P)G^{>}_{ki}(Q)]
  \nn\\
  & \hspace{5mm}+2ip^0\hat{q}^i\hat{q}^jG^{11}_{ij}(P)- 2i q^0\hat{p}^i\hat{p}^jG^{11}_{ij}(Q)
  \bigg\},
  \label{diffstart}
\end{align} 
so that the structure of the time integrations simplifies greatly; hence our choice of introducing the $\int_{t}^{\infty}dt'$
contribution.
Upon using $G_{ij}(P)=(\delta_{ij}-\hat{p}_i\hat{p}_j)G_T(P)$ 
we find
\begin{align}
  \Delta\gamma
  =& - \frac{2g^{4}\CF \Nc}{3}  \Imp 
  \int_Q\int_P  \frac{2\pi\delta(q^0+p^0) }{(\vec p+\vec q)^2}\nn\\
  &\hspace{-1cm}\times\bigg\{
  q_0^2[G^{11}_{T}(P)G^{11}_{T}(Q)-G^{>}_{T}(P)G^{>}_{T}(Q)][d-2+(\hat{p}\cdot\hat{ q})^2]
  \nn\\
  & +i[G^{11}_{T}(P)+ G^{11}_{T}(Q)][1-(\hat{p}\cdot\hat{ q})^2]
  \bigg\}.
  \label{diffintermediate}
\end{align}
$G_T^>(Q)=(\theta(q^0)+n_\mathrm{B}(|q^0|))2\pi\delta(Q^2)$ is purely real and thus does not contribute to $\Delta \gamma$.
$G^{11}_T(Q)=i \mathbf{P} 1/(q_0^2-q^2)+(1/2+n_\mathrm{B}(|q^0|))2\pi\delta(Q^2)$ has both real and imaginary parts,
with $\mathbf{P}$ a principal-value prescription and $n_\mathrm{B}$ the Bose--Einstein distribution, so that
\begin{align}
  \Delta\gamma
  =& -4\frac{g^{4}\CF \Nc}{3}  
  \int_Q\int_p  \frac{2\pi\delta(Q^2)  }{(\vec p+\vec q)^2}\left[\frac12 + n_\mathrm{B}(|q^0|)\right]
  \nn\\
  &\times
  \bigg\{
  \mathbf{P}\frac{q_0^2}{q_0^2-p^2} \big[d-2+(\hat{p}\cdot\hat{ q})^2\big]
  + 1-(\hat{p}\cdot\hat{ q})^2
  \bigg\}\nn\\
  =&\frac83 \als^2\CF\Nc\,\zeta(3)T^3\,,
  \label{difffinal}
\end{align}
where we have also used the $p\leftrightarrow q$ symmetry of the integrand. 
The final integration has been carried out in DR,
showing that Eq.~\eqref{difffinal} is equal to the difference between Eqs.~\eqref{loresult} and \eqref{loresultadj}, as we set 
out to prove.

\bibliographystyle{apsrev4-1}
\bibliography{lit}

\end{document}